\documentclass[10pt,conference]{IEEEtran}
\usepackage{graphics}
\newtheorem{theorem}{Theorem}[section]
\newtheorem{lemma}{Lemma}[section]
\newtheorem{remark}{Remark}[section]
\newtheorem{definition}{Definition}[section]

\begin{document}

\title{Pseudocodeword weights for non-binary LDPC codes}

\author{\authorblockN{Christine A. Kelley}
\authorblockA{Department of Mathematics\\
University of Notre Dame\\
Notre Dame, IN 46556, USA.\\
Email: ckelley1@nd.edu}
\and
\authorblockN{Deepak Sridhara and Joachim Rosenthal}
\authorblockA{
Institut f\"ur Mathematik \\
Universit\"at Z\"urich\\
CH8057 Z\"urich, Switzerland.\\
Email: \{sridhara,rosen\}@math.unizh.ch}}
%

\maketitle

\begin{abstract} Pseudocodewords of $q$-ary LDPC codes are examined and the weight of a 
pseudocodeword on the $q$-ary symmetric channel is defined. The weight definition of a 
pseudocodeword on the AWGN channel is also extended to two-dimensional $q$-ary modulation 
such as $q$-PAM and $q$-PSK. The tree-based lower bounds on the minimum pseudocodeword 
weight are shown to also hold for $q$-ary LDPC codes on these channels. 
\end{abstract}

\section{Introduction}

Low density parity check (LDPC) codes have been shown to achieve near-capacity performance 
over several communication channels. Typically, they are binary linear codes described by 
sparse, randomly, generated parity-check matrices. In \cite{da98} and \cite{sr05}, the 
performance of non-binary LDPC codes, defined over larger finite fields and over integer 
rings, is investigated and compared with that of binary LDPC codes. For several 
applications such as coded-modulation, codes over higher alphabets are more appropriate 
for system design. The popularity of LDPC codes is due to their efficient and simple 
decoding. Graph-based message passing iterative decoders have been shown to achieve 
near-capacity performance with complexity only linear in the length of the code. However, 
these iterative decoders are sub-optimal and discrepancies between iterative and 
maximum-likelihood (ML) decoding performance of short to moderate block length binary LDPC 
codes has been attributed to the presence of pseudocodewords of the LDPC constraint graphs 
(or, Tanner graphs) \cite{ko03p}. Analogous to the role of minimum Hamming distance, 
$d_{\min}$, in ML-decoding, the minimum pseudocodeword weight, $w_{\min}$, has been shown 
to be a leading predictor of performance in iterative decoding \cite{ko03p}.  
Furthermore, it has been observed that pseudocodewords with weight $w_{\min} < d_{\min}$ 
are especially problematic for iterative decoding \cite{ke05u}. In this paper, we define 
pseudocodeword weights for $q$-ary LDPC codes when the channel is a AWGN channel or a 
$q$-ary symmetric channel and obtain lower bounds for the minimum pseudocodeword weight.

The following section shows a tree-based lower bound on the minimum pseudocodeword weight 
of binary LDPC codes. In Section III, the pseudocodeword weight of $q$-ary LDPC codes is 
defined for the AWGN and the $q$-ary symmetric channels. Subsequently, the tree-based 
lower bound for binary LDPC codes is extended to the $q$-ary setting.  We note here that 
we restrict our analysis to pseudocodewords arising from finite-degree graph covers as 
described in \cite{ko03p}. Since these pseudocodewords are the same as those occurring in 
the context of linear programming (LP) decoding, the results obtained here are applicable 
to pseudocodewords of LP decoding as well. Section IV summarizes the paper and outlines 
some other techniques that are being investigated for bounding the pseudocodeword weight 
of $q$-ary LDPC codes.

\section{Binary LDPC codes}

\begin{definition}
The tree bound of a $d$ left (variable node) regular bipartite
LDPC constraint graph with girth $g$ is defined as 

{\tiny \begin{equation}
T(d,g):= \left\{\begin{array}{cc}
1+d + d(d-1) + d(d-1)^2+\ldots+ d(d-1)^{\frac{g-6}{4}}, & 
\frac{g}{2}\mbox{ odd },\\
1+ d+d(d-1)+\ldots+d(d-1)^{\frac{g-8}{4}} + (d-1)^{\frac{g-4}{4}},& 
\frac{g}{2}\mbox{ even }.
\end{array}\right . 
\end{equation}
}
\label{treebound_defn}
\end{definition}

\begin{theorem}
{ \em Let $G$ be a bipartite LDPC constraint graph with smallest left 
(variable node) degree $d$ and girth $g$. Then the minimum pseudocodeword 
weight $w_{\min}$ is lower bounded by
\vspace{-0in}
 \[\scriptsize w_{\min} \ge T(d,g). \]
on the additive white Gaussian noise (AWGN) channel and the binary symmetric channel (BSC).
}
\label{thm1}
\end{theorem}

The proof of this result is presented in \cite{ke05u}. The tree bound was originally derived by Tanner in \cite{ta81a} to
lower-bound the 
 minimum distance of the code. Since the set 
of pseudocodewords includes all codewords, we have $w_{\min}\le d_{\min}$.
 
\section{Non-binary LDPC codes} 

Let $H$ be a parity check matrix representing a $q$-ary LDPC code $\mathcal{C}$. Thus, $H$ 
is sparse in the number of non-zero entries. The corresponding LDPC constraint graph $G$ 
that represents $H$ is an incidence graph of the parity check matrix as in the binary 
case. However, each edge of $G$ is now assigned a weight which is the value of the 
corresponding non-zero entry in $H$. (In \cite{da98,da99t}, LDPC codes over $GF(q)$ are 
considered for transmission over binary modulated channels, whereas in \cite{sr05}, LDPC 
codes over integer rings are considered for higher-order modulation signal sets.) For 
convenience, we consider the special case wherein each of these edge weights are equal to 
one. This is the case when the parity check matrix has only zeros and ones. Furthermore, 
whenever the LDPC graphs have edge weights of unity for all the edges, we refer to such a 
graph as a binary LDPC constraint graph representing a $q$-ary LDPC code $\mathcal{C}$.

\subsection{Bound on minimum distance}
We first show that if the LDPC graph corresponding to $H$ is $d$-left 
(variable-node) regular, then the same tree bound of Theorem~\ref{thm1} 
holds. That is, \\

\begin{lemma} {\em If $G$ is a $d$-left regular bipartite 
LDPC constraint graph with unity edge weights, girth $g$, and represents a $q$-ary LDPC code 
$\mathcal{C}$. Then the minimum distance of the $q$-ary LDPC code 
$\mathcal{C}$ is lower bounded as
 \[d_{\min} \ge T(d,g). \]
}
\label{q_ary_dmin}
\end{lemma}
\begin{proof}
The proof is essentially the same as in the binary
case. Enumerate the graph as a tree starting at an arbitrary
variable node. Furthermore, assume that a codeword in
$\mathcal{C}$ contains the root node in its support. The root
variable node (at layer $L_0$ of the tree) connects to $d$
constraint nodes in the next layer (layer $L_1$)
of the tree. These constraint nodes are each connected
to some sets of variable nodes in layer $L_2$, and so on. Since
the graph has girth $g$, the nodes enumerated up to layer
$L_{\frac{g-2}{2}}$ when $\frac{g}{2}$ is odd (respectively,
$L_{\frac{g}{2}}$ when $\frac{g}{2}$ is even) are all
distinct. Since the root node belongs to a codeword, say ${\bf c}$, it assumes a
non-zero value in ${\bf c}$. Since the constraints must be
satisfied at the nodes in layer $L_1$, at least one node in Layer
$L_2$ for each constraint node in $L_1$ must assume a
non-zero value in ${\bf c}$. (This is true under the assumption that
an edge weight 
times a (non-zero) value, assigned to the corresponding variable node, is non-zero in 
the code alphabet.)

Under the above assumption, there are at least $d$ variable nodes (i.e.,
at least one for each node in layer $L_1$) in layer $L_2$ that are
non-zero in ${\bf c}$. Continuing this argument, it is easy to see that
the number of non-zero components in ${\bf c}$ is at least
$1+d+d(d-1)+\dots+d(d-1)^{\frac{g-6}{4}}$ when $\frac{g}{2}$ is odd, and
$1+d+d(d-1)+\dots+d(d-1)^{\frac{g-8}{4}} +(d-1)^{\frac{g-4}{4}}$ when
$\frac{g}{2}$ is even. This proves the desired lower bound. \end{proof}
\vspace{0.1in}

\begin{remark} A non-zero edge-weight times a (non-zero) value, assigned
to the corresponding variable node, may be zero in certain code alphabets.
Since we have chosen the edge weights to be unity, such a case will not
arise here. But also more generally, such cases will not arise when the
alphabet and the arithmetic operations correspond to finite-field
operations. However, when working over other structures, such as finite
integer rings and more general groups, such cases could arise.  
\end{remark} \vspace{0.1in}

We note here that in general this lower bound is not met and typically
$q$-ary LDPC codes that have the above graph representation have minimum
distances larger than the above lower bound.

\subsection{Pseudocodewords of $q$-ary LDPC codes}

Recall from \cite{ko03p,ke05u} that a pseudocodeword of an LDPC constraint
graph $G$ is a valid codeword in some finite cover of $G$. To define a
pseudocodeword for a $q$-ary LDPC code, we will restrict the discussion to
LDPC constraint graphs that have edge weights of unity among all their
edges -- in other words, binary LDPC constraint graphs that represent
$q$-ary LDPC codes.  A finite cover of a graph is defined in a natural way
as in \cite{ko03p} wherein all edges in the finite cover also have an edge
weight of unity. For the rest of this section, let $G$ be a LDPC
constraint graph of a $q$-ary LDPC code $\mathcal{C}$ of block length $n$,
and let the weights on every edge of $G$ be unity. We define a
pseudocodeword $F$ of $G$ as a $n\times q$ matrix of the form
\[ F= \left[\begin{array}{ccccc} 
f_{0,0}&f_{0,1}& f_{0,2}&\dots&f_{0,q-1}\\
f_{1,0}&f_{1,1}& f_{1,2}&\dots&f_{1,q-1}\\
\vdots&\vdots&\vdots&\vdots&\vdots\\
f_{n-1,0}&f_{n-1,1}& f_{n-1,2}&\dots&f_{n-1,q-1}
\end{array}\right] , \]
where the pseudocodeword $F$ forms a valid codeword
$\hat{\bf c}$ in a finite cover $\hat{G}$ of $G$ and $f_{i,j}$ is
the fraction of variable nodes in the $i^{th}$ variable node cloud, for
$0\le i\le n-1$, of
$\hat{G}$ that have the assignment (or, value) equal to $j$, for
$0\le j\le q-1$, in $\hat{\bf c}$. 

\begin{figure}
\centering{\resizebox{2.5in}{2.3in}{\includegraphics{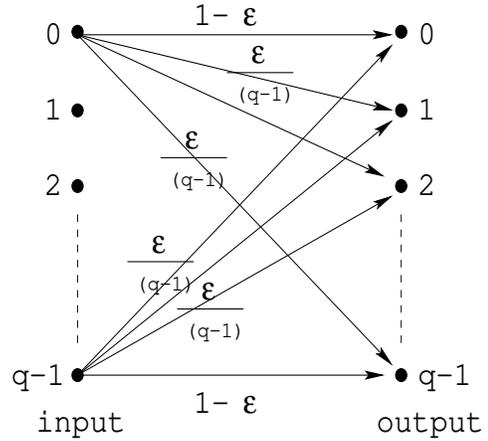}}}
\caption{A $q$-ary symmetric channel.}
\label{q_ary_sym}
\end{figure}

A $q$-ary symmetric channel is shown in Figure~\ref{q_ary_sym}. The input
and the output of the channel are random variables belonging to a $q$-ary
alphabet that can be denoted as $\{0,1,2,\dots,q-1\}$. An error occurs
with probability $\epsilon$, which is parameterized by the channel, and in
the case of an error, it is equally probable for an input symbol to be
altered to any one of the remaining symbols.

Following the definition of pseudocodeword weight for the binary symmetric channel \cite{fo01in}, 
we provide the following definition for the weight of a pseudocodeword on the $q$-ary 
symmetric channel. For a pseudocodeword $ F$, let $ F'$ be the sub-matrix obtained by 
removing the first column in $F$. (Note that the first column in $F$ contains the entries 
$f_{0,0}, f_{1,0}, f_{2,0}, \dots,f_{n-1,0}$.) Then the weight of a pseudocodeword $F$ on 
the $q$-ary symmetric channel is defined as follows. \vspace{0.1in}

\begin{definition} Let $e$ be the smallest number such that the sum of the $e$ largest components in
the matrix $F'$, say,
$f_{i_1, j_1},f_{i_2, j_2},\dots,f_{i_e,j_e}$, exceeds $\sum_{i\ne i_1,i_2,\dots,i_e}(1-f_{i,0})$.
Then the weight of $F$ on the $q$-ary symmetric channel is
defined as 
{\tiny \[ w_{qSC}(F) =\left\{ \begin{array}{cc}
           2e, & \mbox{if } f_{i_1,j_1}+\dots +f_{i_e,j_e} =\sum_{i\ne i_1,i_2,\dots,i_e}(1-f_{i,0}),\\
2e-1, & \mbox{if } f_{i_1,j_1}+\dots +f_{i_e,j_e} > \sum_{i\ne i_1,i_2,\dots,i_e}(1-f_{i,0}).
\end{array} \right  . 
\] 
}
\end{definition}
\vspace{0.1in}

Note that in the above definition, none of the $j_k$'s, for $k=1,2,\dots,e$, are equal to 
zero, and all the $i_k$'s, for $k=1,2,\dots,e$, are distinct. That is, we choose at most 
one component from every row of $F'$ when choosing the $e$ largest components. The 
following sub-section provides an explanation for the above definition of weight.

\subsection{\sc Pseudocodeword weight for $q$-ary LDPC codes on
    the $q$-ary symmetric channel}

Suppose the all-zero codeword is sent across a $q$-ary symmetric channel and the vector 
${\bf r}=(r_0,r_1,\dots,r_{n-1})$ is received. Then errors occur in positions where 
$r_i\ne 0$. Let $S=\{i |\ r_i\ne 0\}$ and let $S^c=\{i |\ r_i=0\}$. The distance between 
${\bf r}$ and a pseudocodeword $F$ is defined as
\begin{equation}
d({\bf r},F)=\sum_{i=0}^{n-1}\sum_{k=0}^{q-1}\chi(r_i\ne
k) f_{i,k}, \label{dist}\end{equation}
where $\chi(P)$ is an indicator function that is equal to $1$
if the proposition $P$ is true and is equal to $0$ otherwise.

The distance between ${\bf r}$ and the all-zero codeword ${\bf 0}$ is \[d({\bf r},{\bf 
0})=\sum_{i=0}^{n-1}\chi(r_i\ne 0) \] which is the Hamming weight of ${\bf r}$ and can be 
obtained from equation (\ref{dist}).

The iterative decoder chooses in favor of $F$ instead of the
all-zero codeword ${\bf 0}$ when $d({\bf r},F)\le d({\bf r},{\bf  0})$.
 That is, if 
\[ \sum_{i\in S^c}(1-f_{i,0}) +\sum_{i\in S}(1-f_{i,r_i})\le
\sum_{i\in S}1\]
The condition for choosing $F$ over the all-zero codeword reduces
to 
\[ \Big{\{} \sum_{i\in S^c}(1-f_{i,0})\le \sum_{i\in
  S}f_{i,r_i}\Big{\}}\]
 Hence, we define the weight of a pseudocodeword $F$ in the
 following manner.

 Let $e$ be the smallest number such that the sum of the $e$ largest components in
the matrix $F'$, say,
$f_{i_1,j_1},f_{i_2,j_2},\dots,f_{i_e,j_e}$, exceeds $\sum_{i\ne i_1,i_2,\dots,i_e}(1-f_{i,0})$.
Then the weight of $F$ on the $p$-ary symmetric channel is
defined as 
{\tiny \[ w_{qSC}(F) =\left\{ \begin{array}{cc}
           2e, & \mbox{if } f_{i_1,j_1}+\dots +f_{i_e,j_e} =\sum_{i\ne i_1,i_2,\dots,i_e}(1-f_{i,0})\\
2e-1, & \mbox{if } f_{i_1,j_1}+\dots +f_{i_e,j_e} > \sum_{i\ne i_1,i_2,\dots,i_e}(1-f_{i,0})
\end{array} \right  . 
\] }
\vspace{0.1in}

Note that in the above definition, none of the $j_k$'s, for $k=1,2,\dots,e$, are equal to 
zero, and all the $i_k$'s, for $k=1,2,\dots,e$, are distinct. That is, we choose at most 
one component in every row of $F'$ when picking the $e$ largest components. The received 
vector ${\bf r}=(r_0,r_1,\dots,r_{n-1})$ that has the following components: $r_{i_1}=j_1, 
r_{i_2}=j_2, \dots, r_{i_e}=j_e$, $r_i=0$, for $i\notin \{i_1,i_2,\dots,i_e\}$, will cause 
the decoder to make an error and choose $F$ over the all-zero codeword.

Observe that for a codeword, the above weight definition reduces to the Hamming weight. If 
$F$ represents a codeword ${\bf c}$, then exactly $w=wt_H({\bf c})$, the Hamming weight of 
${\bf c}$, rows in $F'$ contain the entry $1$ in some column, and the remaining entries in 
$F'$ are zero. Furthermore, the matrix $F$ has the entry $0$ in the first column of these 
$w$ rows and has the entry $1$ in the first column of the remaining rows. Therefore, from 
the weight definition of $F$, $e=\frac{w}{2}$ and the weight of $F$ is $2e=w$.

\subsection{\sc Tree Bound on the $q$-ary symmetric channel}
We define the $q$-ary minimum pseudocodeword weight of $G$ (or, minimum
pseudoweight) as in the binary case, i.e., as the minimum weight
of a pseudocodeword among all finite covers of $G$, and denote
this as $w_{\min}(G)$ or $w_{\min}$ when it is clear that we are
referring to the graph $G$. 
\vspace{0.1in}

\begin{theorem}
{\em Let $G$ be a $d$-left regular bipartite graph with girth $g$ that
represents a $q$-ary LDPC code $\mathcal{C}$. Then the minimum
pseudocodeword weight 
$w_{\min}$ on the $q$-ary symmetric channel is lower bounded as
\[w_{\min} \ge T(d,g)\]
}
\label{q_ary_wmin}
\end{theorem}
\vspace{-0.1in}

\begin{proof} 
{\begin{center}
\begin{figure}[h]
\centering{
            \begin{minipage}[b]{0.35\linewidth} 
            \centering
                 {
             \resizebox{1in}{1in}{\vspace{1in}\includegraphics{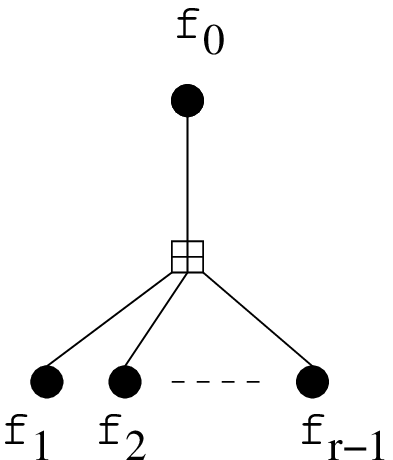}\vspace{0in}}
             \caption{Single constraint code.}\label{spc}
\vspace{0.2in}
             $({1-f_{i,0}})\le \sum_{j\ne i}(1-f_{j,0})$\vspace{0.1in}}
         \end{minipage}
            \hspace{0.1in}
      \begin{minipage}[b]{0.6\linewidth}\vspace{0in}
    \centering{
            \resizebox{1.8in}{1.8in}{\includegraphics{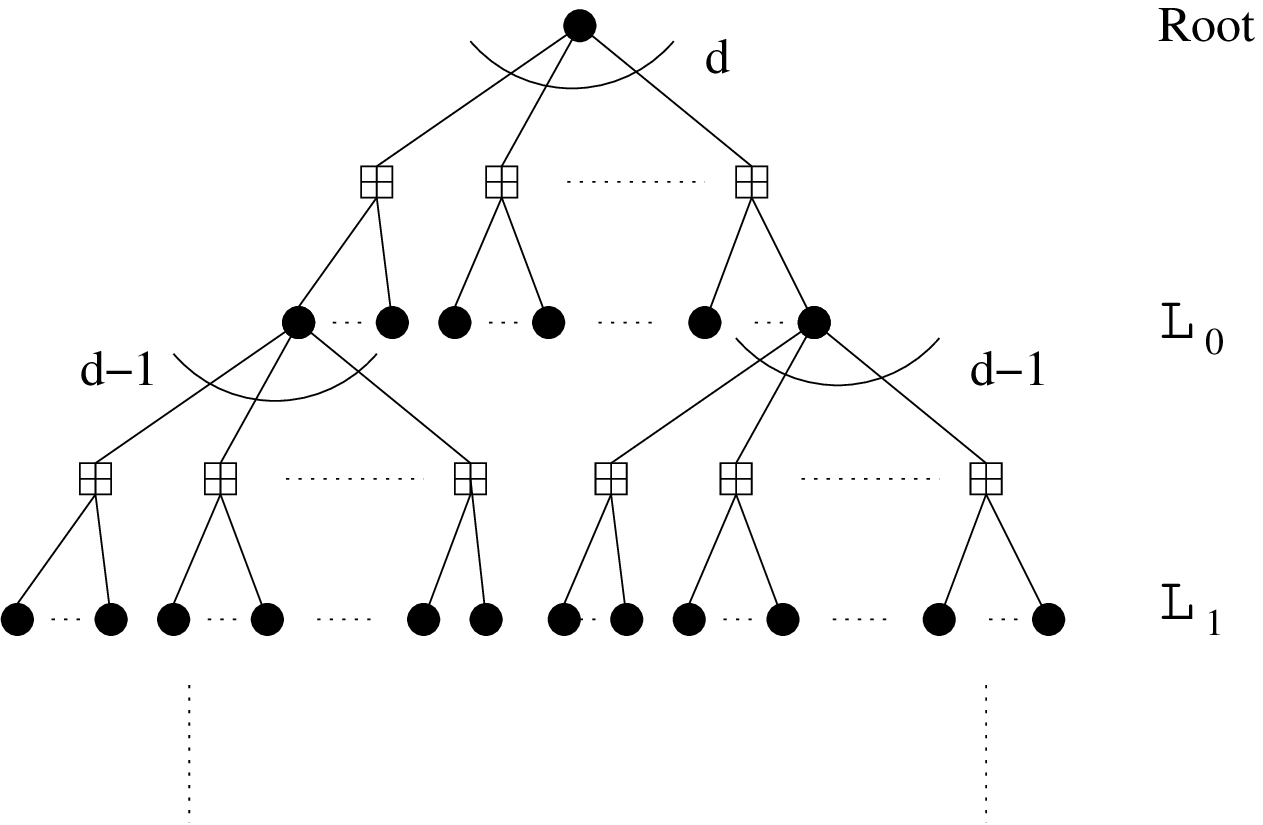}}}
            \caption{Local tree structure for a $d$-left regular graph.}
        \label{tree}
$d (1-f_{0,0}) \le \sum_{j\in
          L_0}(1-f_{j,0})$,\vspace{-0.0in}\\ 
 $d(d-1)(1-f_{0,0})\le \sum_{j\in L_1}(1-f_{j,0})$\vspace{-0.05in}\\
   $:$\vspace{-0.0in}\\
           \end{minipage}}
\end{figure}
\end{center}
}

\underline{Case:} $\frac{g}{2}$ odd. Consider a single constraint node
with $r$ variable node neighbors as shown in
Figure~\ref{spc}. Then, for $i=0,1,\dots, r-1$ and $k=0,1,\dots,p-1$,
the following inequality holds: 
%
\begin{eqnarray}
(1-f_{i,0})\le   \sum_{j\ne i} (1-f_{j,0})
\label{eqn_spc}
\end{eqnarray}

Now let us consider a $d$-left regular LDPC constraint graph representing a $q$-ary LDPC 
code. We will enumerate the LDPC constraint graph as a tree from an arbitrary root 
variable node, as shown in Figure~\ref{tree}. Let $F$ be a pseudocodeword matrix for this 
graph. Without loss of generality, let us assume that the component $(1-f_{0,0})$ 
corresponding to the root node is the maximum among all $(1-f_{i,0})$ over all $i$.

Applying the inequality in (\ref{eqn_spc}) at every constraint node in first constraint 
node layer of the tree, we obtain

\[ d(1-f_{0,0}) \le \sum_{j\in L_0}(1-f_{j,0}),\] where $L_0$
corresponds to variable nodes in first level of the
tree. Subsequent application of the inequality in (\ref{eqn_spc})
to the second layer of constraint nodes in the tree yields
 \[  d(d-1)(1-f_{0,0})\le \sum_{j\in L_1}(1-f_{j,0}),\]
 Continuing this process  until layer $L_{\frac{g-6}{4}}$,
we obtain 
\[d(d-1)^{\frac{g-6}{4}}(1-f_{0,0})\le \sum_{j\in  L_{\frac{g-6}{4}}}(1-f_{j,0})\]

Since the LDPC graph has girth $g$, the variable nodes up to level
$L_{\frac{g-6}{4}}$ are all distinct. The above
inequalities yield:
    \begin{eqnarray}
[1+d+d(d-1)+\dots+d(d-1)^{\frac{g-6}{4}}](1-f_{0,0})  \nonumber \\
\le \sum_{i\in\{0\}\cup L_0\cup\dots L_{\frac{g-6}{4}}} (1-f_{i,0})\le
\sum_{\mbox{ all }i}(1-f_{i,0})
\label{fin_ineq}
\end{eqnarray} 

Let $e$ the smallest number such that there are $e$ maximal
components  $f_{i_1,j_1}$, $f_{i_2,j_2}, f_{i_3,j_3},\dots,
f_{i_e,j_e}$, for $i_1,i_2,\dots,i_e$ all distinct and
$j_1,j_2,\dots,j_e \in \{1,2,\dots,q-1\}$,  in $F'$ (the sub-matrix of $F$
excluding the first column in $F$) such that 
\[f_{i_1,j_1}+f_{i_2,j_2}+\dots+f_{i_e,j_e}\ge \sum_{i\notin
  \{i_1,i_2,i_3,\dots,i_e\}}(1-f_{i,0}) \]

Then, since none of the $j_k$'s, $k=1,2,\dots,e$, are zero, we have
\[ (1-f_{i_1,0})+(1-f_{i_2,0})+\dots+(1-f_{i_e,0})\ge f_{i_1,j_1}+\dots+f_{i_e,j_e}
\]\[\ge \sum_{i\notin
  \{i_1,i_2,i_3,\dots,i_e\}}(1-f_{i,0})\]
Hence we have that
\[2((1-f_{i_1,0})+(1-f_{i_2,0})+\dots+(1-f_{i_e,0})) \]\[\ge \sum_{\mbox{all }
  i}(1-f_{i,0})\]

We can then lower bound this further using the inequality in
(\ref{fin_ineq}) as 
\[2((1-f_{i_1,0})+(1-f_{i_2,0})+\dots+(1-f_{i_e,0}))\]\[\ge
[1+d+d(d-1)+\dots+d(d-1)^{\frac{g-6}{4}}](1-f_{0,0})\]
Since we assumed that $(1-f_{0,0})$ is the maximum among 
$(1-f_{i,0})$ over all $i$,   we have
\[2e(1-f_{0,0})\ge 2((1-f_{i_1,0})+(1-f_{i_2,0})+\dots+(1-f_{i_e,0}))\]\[\ge
[1+d+d(d-1)+\dots+d(d-1)^{\frac{g-6}{4}}](1-f_{0,0})\]
This yields the desired bound \[w_{qSC}(F)= 2e\ge 
1+d+d(d-1)+\dots+d(d-1)^{\frac{g-6}{4}}.\]
Since the pseudocodeword $F$ was chosen arbitrary, we also have
$w_{\min}\ge 1+d+d(d-1)+\dots+d(d-1)^{\frac{g-6}{4}}$. The case
$\frac{g}{2}$ even is treated similarly. 
\end{proof}
\vspace{0.1in}
          
Since the inequality in
(\ref{eqn_spc}), in the proof of Theorem~\ref{q_ary_wmin}, is
typically not tight, the above bound is rather loose.  

\subsection{\sc Pseudocodeword weight on the AWGN channel}
Following the definition of effective distance $d_{eff}^2({ F},{\bf c})$, 
between a pseudocodeword ${F}$ and a codeword ${\bf c}$ on the AWGN 
channel,  presented in \cite{fo01in}, the 
weight of a pseudocodeword ${ F}$
is given by $d_{eff}^2({ F},{\bf 0})$. On simplifying the expression 
in \cite{fo01in}, the weight of pseudocodeword ${F}$ on the AWGN 
channel is given by 
\[ w_{q-AWGN}({ F})=\frac{(\sum_{i=0}^{n-1}\sum_{m=0}^{q-1} f_{i,m} 
m^2)^2}{\sum_{i=0}^{n-1}(\sum_{m=0}^{q-1} f_{i,m} m)^2} \ \ (*)
\]
The above weight definition assumes $q$-ary pulse amplitude modulation, i.e., the symbols sent 
across the channel belong to the signal set $\{0,1,2,\dots,q-1\}$.

Now if we assume a two-dimensional signal set for transmission on the 
memoryless AWGN 
channel, then under the assumption that the resulting signal-space code 
is geometrically uniform 
\cite{forney2}, we 
can derive the weight of a
pseudocodeword $F$ as the effective distance of $F$ from the all-zero codeword in signal 
space. The pseudocodeword weight of $F$ is given by
\[ w_{q-AWGN}(F) = \frac{(R-M)^2}{V},\]
where $(x_m, y_m)$ is the coordinate in the two-dimensional signal set corresponding to 
the symbol $m \in \{0,1,\ldots,q-1\}$, \[R =\sum_j [\sum_m f_{j,m} (x_m^2+y_m^2)-x_0^2-y_0^2],\]
\[M = 2\sum_j[(\sum_m f_{j,m} x_m x_0)-x_0^2 + (\sum_m f_{j,m}y_m y_0)-y_0^2],\] 
 \[V = 4\sum_j [((\sum_m f_{j,m} x_m) - x_0)^2 + \sum_j ((\sum_m f_{j,m}y_m) - y_0)^2 
],\] and $j \in
\{0,\ldots n-1\}$.

Note that for $q$-ary pulse amplitude modulation as described above, this weight 
definition reduces to the one in $(*)$.

Suppose we assume $q$-PSK modulation, then we have $x_m =
\cos(\frac{2\pi m}{q})$ and $y_m = \sin(\frac{2 \pi m}{q})$. Note that $x_0 = \cos(0) = 1$ 
and $y_0 =
\sin(0) = 0$. In addition, $R = 0$. Therefore, the weight of a pseudocodeword $F$ on the 
AWGN channel under $q$-PSK modulation
is given by: $w_{q-AWGN}(F) =\frac{M^2}{V}$,
where
{\scriptsize \[M= 2\sum_j((\sum_m f_{j,m} \cos(\frac{2\pi
m}{q}))-1)\]
\[V= 4\sum_j \Big{[} \sum_m f_{j,m}^2 + 2(\sum_{m,m';m\ne m'} f_{j,m} 
f_{j,m'} 
(\cos(\frac{2 
\pi
(m-m')}{q})))\]\[ -2\sum_m f_{j,m}\cos(\frac{2 \pi m}{q}) + 1  
\Big{]}  .\]
}


\subsection{\sc Tree-bound of $q$-ary LDPC codes on the AWGN channel under $q$-PAM}
\vspace{0.1in}
\begin{theorem}[$q$-ary pulse amplitude modulation]
{\em Let $G$ be a $d$-left regular bipartite graph with girth $g$ that
represents a $q$-ary LDPC code $\mathcal{C}$. Then the minimum
pseudocodeword weight 
$w_{\min}$ on the AWGN channel is lower bounded as

\[w_{\min} \ge T(d,g).\]
}
\label{q_ary_awgn_wmin}
\end{theorem}
\vspace{0.1in}

(Note that we assume a slightly unconventional definition of $q$-ary PAM in that the 
symbol $m$ is mapped to the point $m$ rather than to the point $2m-1$ as in the 
conventional definition, for $m\in \{0,1,2,\dots,q-1\}$.)
 
\begin{proof}
Let ${ F}$ be a pseudocodeword in $G$.
Without loss of generality, let $(1-f_{0,0})$ be the maximum of $(1-f_{0,i})$ over all $i$.
We will first lower bound the weight $w_{q-AWGN}({ F})$ as 
\[ w_{q-AWGN}({ F}) = \frac{(\sum_{i=0}^{n-1}\sum_{m=0}^{q-1} f_{i,m} m^2)^2}{\sum_{i=0}^{n-1}(\sum_{m=0}^{q-1} f_{i,m} m)^2}
\]\[\ge \frac{(\sum_{i=0}^{n-1}\sum_{m=0}^{q-1}f_{i,m}m^2)}{1-f_{0,0}} \ \ (**)\] 

This lower bound is obtained by showing that the denominator in the weight expression
can be upper bounded by using the Cauchy-Schwartz inequality as follows \[\sum_{i=0}^{n-1}(\sum_{m=0}^{q-1}f_{i,m}m)^2 \]
\[\le (\sum_{i=0}^{n-1}(f_{i,1}+f_{i,2}+\dots+ f_{i,q-1}))(\sum_{i=0}^{n-1}\sum_{m=0}^{q-1}f_{i,m}m^2).\]
Further, since  $f_{i,1}+f_{i,2}+\dots+f_{i,q-1}=1-f_{i,0}\le 1-f_{0,0}$, we obtain the lower bound
in $(**)$.\\

Since $\sum_{i=0}^{n-1}\sum_{m=0}^{q-1}f_{i,m}m^2 \ge \sum_{i=0}^{n-1}(f_{i,1}+\dots+f_{i,q-1}) =
\sum_{i=0}^{n-1}(1-f_{i,0})$, we have \[w_{q-AWGN}({ F}) \ge \frac{\sum_{i=0}^{n-1}(1-f_{i,0})}{1-f_{0,0}}\]
Now, the inequality (\ref{fin_ineq}) from the proof of Theorem~\ref{q_ary_wmin} yields the desired lower bound
$w_{q-AWGN}({ F}) \ge 1+d+d(d-1)+\dots+d(d-1)^{\frac{g-6}{4}}$ for the case $g/2$ odd.
(The case $g/2$ even follows similarly.)
 \end{proof}
\vspace{0.1in}


\section{Conclusions} This paper examined the pseudocodeword weight of
$q$-ary LDPC codes on the $q$-ary symmetric channel and the AWGN channel.  
A definition for the pseudocodeword weight was derived on the $q$-ary
symmetric channel and the AWGN channel with two-dimensional $q$-ary
modulation. The tree bound from \cite{ke05u} for binary LDPC codes was
extended to the $q$-ary case. More sophisticated bounding techniques for
the pseudocodeword weight of $q$-ary LDPC codes remains an open problem.
It would be useful to also derive a cost-function of the min-sum decoder
for $q$-ary LDPC codes to give an insight into which pseudocodewords are
problematic for iterative decoding.



\section*{Acknowledgment} 
This work was supported in part by the NSF Grant No. CCR-ITR-02-05310.

\end{document}